\documentclass[english,keywords,amsmath,amssymb,twocolumn]{revtex4}
\usepackage[T1]{fontenc}
\usepackage[latin1]{inputenc}
\usepackage{babel}
\usepackage{graphicx}
\usepackage{color}
\usepackage{bm}
\usepackage{longtable}
\usepackage{amsmath}
\usepackage{amsfonts}
\usepackage{amssymb}
\usepackage{hyperref}
\begin{document}
\title{Poissonian twin beam states and the effect of symmetrical photon subtraction in loss estimations}
\author{N.~Samantaray}
\email{Email: ns17363@bristol.ac.uk}
\author{J. C. F.~Matthews}
\author{J. G.~Rarity}

\affiliation{Quantum Engineering Technology Labs, H. H. Wills Physics Laboratory and
Department of Electrical and Electronic Engineering, University of Bristol, BS8 1FD, UK}

\begin{abstract}
We have devised an experimentally realizable model generating twin beam states whose individual beam photon statistics are varied from thermal to Poissonian keeping the non-classical mode correlation intact. We have studied the usefulness of these states for loss measurement by considering three different estimators, comparing with the correlated thermal twin beam states generated from spontaneous parametric down conversion or four wavemixing . We then incorporated the photon subtraction operation into the model and demonstrate their advantage in loss estimations with respect to un-subtracted states at both fixed squeezing and per photon exposure of the absorbing sample. For instance, at fixed squeezing, for two photon subtraction, up to three times advantage is found. In the latter case, albeit the advantage due to photon subtraction mostly subsides in standard regime, an unexpected result is that in some operating regimes the photon subtraction scheme can also give up to 20$\%$ advantage over the correlated Poisson beam result. We have also made a comparative study of these estimators for finding the best measurement for loss estimations. We present results for all the values of the model parameters changing the statistics of twin beam states from thermal to Poissonian.
\end{abstract}
\maketitle

\section{Introduction}
Absorption based measurement underpins many approaches to spectroscopy and imaging. It finds application in all branches of science from chemistry and biology to physics and material science. However, the best sensitivity in loss estimation reached so far using classical light probes is limited by photon shot noise. In last years, non classical resources such as non-Gaussian states (by de-Gaussification of Gaussian states) have shown to reach sub-shot noise (SSN) limit in loss estimations in terms of Fisher information \cite{Olivares:2003}. De-Gaussified single mode squeezed vacuum has been reported for theoretical quantum enhancement in loss estimation \cite{Gerardo:2009}. Two basic operations that can lead to non-Gaussian states are photon addition to, or photon subtraction from  Gaussian light states \cite{kim:2008, kim1:2008}. Another important feature of bi-partite quantum states to reach sub-shot noise limit (SSNL) are non-classical correlations \cite{Rarity:1991}. 
 \par It is known that the twin beam state (TBS) generated by spontaneous parametric down conversion (SPDC) or four wave mixing (FWM) process has thermal photon statistics in the individual modes, but its perfect photon number non classical mode correlation allows surpassing the shot-noise limit (SNL) reaching SSN sensitivity in the realistic scenario of loss estimations \cite{Chekova:2015, Brida:2009,Lopaeva:2013, Marco:2017,Moreau:2017,Javier:2019,Javier:2017}. More recently, unbiased estimations of optical losses \cite{Losero:2018} (losses are estimated in an absolute way without pre-calibration of the apparatus) at ultimate quantum limit  have been reported exploiting the quantum correlations in TBS. In the laboratory context, these correlated beams usually appear Poissonian due to temporal (or spatial) averaging of thermal statistics. \par De-Gaussification by symmetrical photon subtraction on both of the mode of TBS has not only been shown to improve the individual mode photon statistics from thermal to sub-Poissonian \cite{chekova:2016}, but it also increases the entanglement between them \cite{Carlos:2012,Tim:2013,ourjoumtsev:2007}. In the last years, the resulting TBS states after photon subtraction have been theoretically investigated reporting their advantage over TBS for target detection in the presence of noise, the so called "quantum illumination" \cite{Zhang:2014}. Their advantage over TBS have also been demonstrated in single interferometry with parity measurements \cite{raul:2012} and more recently for probing the Plank scale physics \cite{Nigam:2019} and distillation of squeezing \cite{Thomas:2019}. Looking at all these advantages of symmetrical photon subtracted TBS (SPSTBS) over TBS because of their improved photon statistics and non-classical correlation, we  proliferated our interest for using SPTBS for loss estimations. However, a question on fundamental grounds naturally arises: does photon subtraction have any advantage in noise suppression if the individual mode photon statistics of TBS becomes Poissonian (due to averaging of thermal statistics). Keeping this motivation in mind, we have devised a theoretical but experimental realizable model (accounting detection losses), where changing the value of a parameter of the models changes the TBS individual mode photon statistics from thermal to Poissonian. \par
\begin{figure*}[htb]
\centering
\setlength\fboxsep{0pt}
\setlength\fboxrule{0.25pt}
\fbox{\includegraphics[width=6.9in]{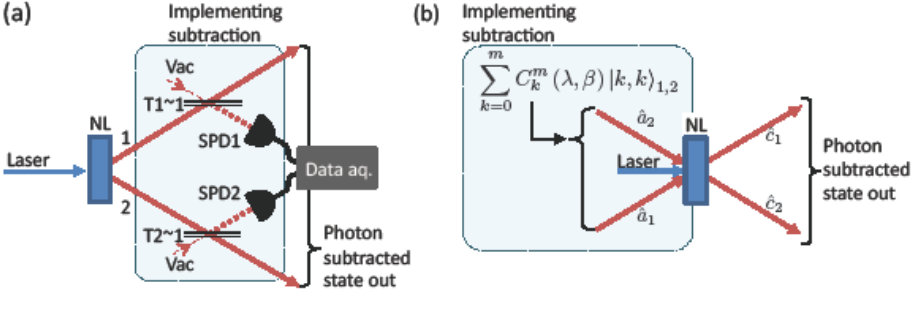}}
\caption{ An equivalent ways of getting photon subtracted state: (a) The left hand side image represents the conventional approach in which two high transmittance beam splitters are placed in the each path of TBS and a simultaneous photon clicks on the single photon detectors confirms the implementation of symmetrical photon subtraction, this subtraction operation can be implemented equivalently by seeding a superposition state to the non-linear crystal (NL) together with the pump beam as shown in the right image (b).}
\label{equivalence}
\end{figure*}

 For a null value of the model parameter, i.e, when the statistics of individual beams are Poissonian, the resulting state becomes correlated Poissonian TBS (CPTBS) keeping the initial TBS non-classical mode correlation intact. We then incorporated the symmetrical photon subtraction into the model of absorption measurement. We replace the conventional approach of obtaining the photon subtraction(placing high transmittance beam splitters on the individual beam paths ) by an alternate way of seeding the photon number super-position state \cite{Nigam:2019} to the squeezer as shown in the Fig.\ref{equivalence}. 
 
 One of the important goals of this letter is to answer the following question. To what extent does photon subtraction bring advantage for absorption measurement compared to TBS at both fixed squeezing and per photon exposure to the absorbing sample and furthermore does photon subtraction provide any advantage particularly when the individual TBS mode statistics turns to Poissonian. Apart from a few recent works which consider phase measurement \cite{Nigam:2019, raul:2012}, most of the quantum optics and information protocols demonstrate the advantage of photon subtraction at a fixed squeezing parameter \cite{Olivares:2003,Genoni:2010,Optarny:2000,Cochrane:2002}. \par This paper is organized in the following way. In section \ref{Absorption}, we shall briefly describe the absorption measurement and various types of estimators for loss estimation. Importance of photon statistics and non-classical correlation in measuring these estimators will also be addressed. Section  \ref{Model1} details this model and a way to incorporate photon subtraction operation. We shall also present results for different types of absorption estimators up to two photon subtraction and discuss the usefulness of our model. All the values of the model parameter that change the statistics of TBS from thermal to Poisonian and in between have also been considered in the result. We conclude the paper with a summary in section \ref{Conclusions}.
\section{Absorption measurement:}\label{Absorption}
\begin{figure}[htb]
	\centering
	\includegraphics[width=7.5cm,height=8cm]{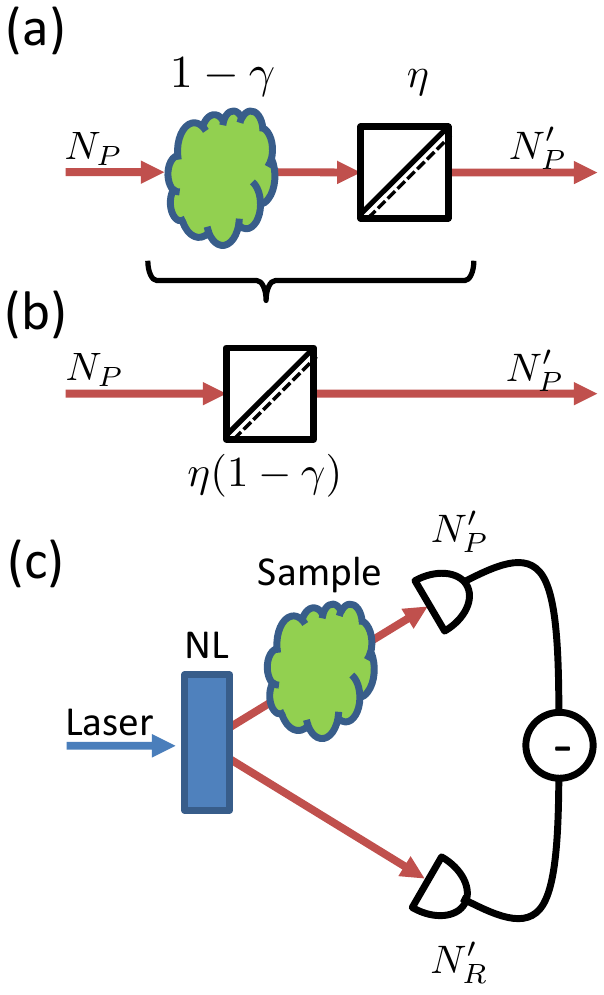}
	\caption{Absorption measurement: (a) direct one path imaging of a sample object of absorption coefficient $\gamma$ and $\eta$ is the detection loss, (b) beam splitter equivalence of reflection and detection losses in direct one path imaging, (c) photon number difference measurement in the presence of quantum correlation generated by pumping the non-linear crystal in SPDC process.}
	\label{scheme}
\end{figure}

Absorption is measured by probing the sample with known light intensity and then measuring the light intensity at the detection stage as shown in the fig.\ref{scheme}, where $\gamma$ is the absorption coefficient, $\eta$ is the detection loss, $N_{P}$ and $N^{'}_{P}$ are the number of detected photons before and after placing the sample respectively.
The losses due to the presence of sample and the detection losses are modelled by using a single beam splitter with transmittance $\tau=\eta\left(1-\gamma\right)$. For applications where low light illumination is required, the uncertainty in measuring $\gamma$ is dominated by photon shot noise (SN). The uncertainty in absorption measurement due to photon shot noise can be improved by considering two beams which are correlated in the photon number basis as in TBS states generated by SPDC process. The first beam passes though the sample, whereas the second beam acts as reference thus partially cancelling the SN going below shot noise limit in realistic scenarios. We shall consider three different absorption estimators including a new one based on ratio measurement, the TBS state as input source, and balanced detection (quantum efficiency $\eta$ remains the same for both beams) throughout this work. The principal point here is to show the dependence of uncertainties in measuring different absorption estimators on parameters that characterize photon statistics and correlation of the input probe states such as Fano factor (F), and noise reduction factor denoted by the symbol $\sigma$, and finding the best absorption measurement for loss estimations. Fano factor is defined as the variance of photon number of a state normalized to its mean value. In terms of statistics, it represents how a state is different from a coherent state. $F>1$, $F=1$, and $F<1$ refers to super-Poissonian, Poissonian and sub-Poissonian statistics of the light state respectively. Analogously, for any bi-partite state, $\sigma$ is defined as variance of the photon number difference normalized to their mean. One can easily check $\sigma<1$ refers to non-classical photon number correlation. It is worth checking the change in these two parameters when individual TBS mode statistics change from thermal (super-Poissonian) to Poissonian. We have devised a phenomenological model in this context and we shall detail it in the next sections.  Furthermore, it is interesting for both fundamental perspectives and applications to see the change in photon statistics and correlation by subtracting photons symmetrically from each mode of TBS, and to investigate up-to what extent the photon subtraction operation is advantageous in this scheme.
\subsection{Number difference measurement:}\label{AbsorptionDiff}
In this measurement, the observable under consideration $\hat{o}=\hat{N_{R}}-\hat{N^{'}_{P}}$ is photon number difference of two beams after placing the object as shown in Fig.\ref{scheme} (b). As per theory of error propagation, the uncertainty in measuring $\gamma$ \cite{Gerry:2005} is 
\begin{equation}\label{UncD}
\Delta\gamma_{Diff}=\frac{\sqrt{\Delta^{2}\hat{o}}}{|\frac{\partial \langle\hat{o}\rangle}{\partial\gamma}|}=\sqrt{\frac{\gamma^{2}[F-1]+\gamma+2\sigma\left(1-\gamma\right)}{\langle N_{P}\rangle}},
\end{equation}
where $\langle N_{P}\rangle=\eta\sinh^{2} r = \eta\lambda$ is the detected mean number of photons per mode of the probe TBS state, r being the squeezing parameter. r carries necessary information about the pump intensity and phase matching function of the SPDC process. It is easy to check for classical states, i.e, $F=1$ and $\sigma=1$, a limit $\Delta\gamma_{Diff}=\sqrt{(2-\gamma)/\langle N_{P}\rangle}$  known as the shot noise limit (SNL) in differential absorption measurement. It is paramount to note for no absorption ($\gamma=0$), this limit is twice the standard shot noise limit in direct one path imaging $1/\sqrt{\langle N_{P}\rangle}$. Each beam carries one unit of shot noise although for $\gamma=1$, standard SNL is reached. It can be checked that for low values of of absorption, i.e, $\gamma << 1$, $F=1$ and $\sigma<1/2$ allows beating SNL, where as for relatively high $\gamma$, the probe state with $\sigma<1/2$ and $F<1$ is required for reaching SSN limit.
\subsection{Optimized balanced absorption estimator:}\label{AbsorptionNew}
In the last couple of years, a different absorption estimator of the following form \cite{Moreau:2017} has been considered
\begin{equation}
\gamma_{Opt}=1-\frac{\hat{N^{'}_{P}-k\Delta N_{R}+\delta E}}{\langle N_{P}\rangle},
\end{equation}
where k is a factor to be experimentally determined in order to minimize the uncertainty and $\delta E$ is a correction factor for making the estimator unbiased. Exploiting theory of error propagation, we worked out the uncertainty of this estimator
\begin{equation}\label{newestimator}
\Delta\gamma_{Opt}=\sqrt{\frac{\gamma\left(1-\gamma\right)}{\langle N_{P}\rangle}+\frac{\left(1-\gamma\right)^{2}\sigma}{\langle N_{P}\rangle}\left(2-\frac{\sigma}{F}\right)}
\end{equation}
A clear advantage of this estimator is seen as it is  $\sqrt{2}$ times advantageous compared to the number difference at low absorption ($\gamma\rightarrow 0$) for $\sigma=1$ and $F=1$. Another interesting point about this estimator is the requirement of lower quantum correlation, i.e  $\sigma<1$ ( $\sigma<1/2$ for the number difference case ) and $F<1$ for reaching SSN limit in absorption measurement. It can be easily checked for both $\sigma=0, F=1$ (perfect photon number correlation and Poissonian individual statistics) and $\sigma=1 ,F=1/2$ (classical photon number correlation and sub-poissonian individual statistics), the uncertainty in eq.\ref{newestimator} simplifies to an ultimate quantum limit (UQL)\cite{Losero:2018},
\begin{equation}\label{UQL}
\Delta\gamma_{Opt}=\sqrt{\frac{\gamma\left(1-\gamma\right)}{\langle N_{P}\rangle}}
\end{equation}
\subsection{Ratio measurement:}
We have considered a new type of estimator based on ratio measurement. In this measurement the observable we consider is $\hat{o(\gamma)}=N^{'}_{P}/N_{R}$, or $\hat{o(\gamma)}=N_{R}/N^{'}_{P}$. Propagating the errors, we obtain the expression of uncertainty in ratio measurement as:
\begin{equation}\label{Ratioestimator}
\Delta\gamma_{Ratio}=\sqrt{\frac{\gamma\left(1-\gamma\right)}{\langle N_{P}\rangle}+\frac{\left(1-\gamma\right)^{2}2\sigma}{\langle N_{P}\rangle}}.
\end{equation}
Unlike other two estimators, it shows dependence of the measured absorption uncertainty only on correlation of the probe state $\sigma$. It can be checked for $\sigma=1/2$ and $\sigma=0$ (perfect quantum correlation), the corresponding uncertainty becomes SNL (direct one path imaging) and UQL (eq.\ref{UQL}) respectively. Thus, similar to the number difference measurement, ratio estimator beats direct one path imaging when less than 50$\%$ of the photon number correlation is lost.
\section{Model for correlated TWB state and symmetrical photon subtraction:}\label{Model1}
\begin{figure}[htb]
	\centering
	\includegraphics[width=8cm,height=3.5cm]{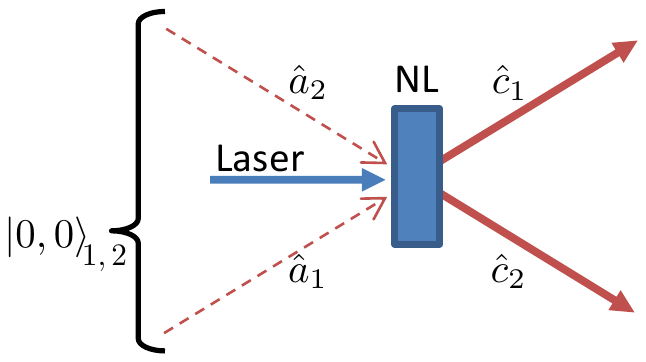}
	\caption{Scheme for generating correlated TBS states: Coherent beam pumps a non linear crystal (NL) for generating correlated twin photon states at the exit face of the crystal.}
	\label{cptwb}
\end{figure}
Considering the quantum picture of the parametric down conversion, we modelled the mode operators before and the after the non-linear crystal (see fig. \ref{cptwb}) in the following way
\begin{equation}\label{c1}
\hat{c_{1}}(\lambda,\beta)\approx\hat{a_{1}}\sqrt{1+\beta\lambda}+\hat{a^{\dagger}_{2}}\sqrt{\lambda}, 
\end{equation}
\begin{equation}\label{c2}
\hat{c_{2}}(\lambda,\beta)\approx\hat{a_{2}}\sqrt{1+\beta\lambda}+\hat{a^{\dagger}_{1}}\sqrt{\lambda},
\end{equation}
where $\beta=1/M$ is the model parameter, $M$ being the number of modes. For $\beta=1$, these set of equations look like the usual Bougolibov transformation when vacuum state turns to two mode squeezed state by an action of two mode squeezed operator $\hat{S}_{1,2}(\lambda)$. In experimental situation, significance of $\beta$ can be seen as follows: for single temporal mode in the time window of picoseconds which is the coherence time $\tau_{coh}$ of the SPDC, $\beta=1$, whereas for more time exposure, many temporal modes are collected and in that case $\beta\approx 0$. The last case is realised for high temporal band width pump beam so that many more modes $M=\tau_{p}/\tau_{coh}$ are generated. This situation is usually considered experimentally for alleviating the excess noise from the individual TBS demonstrating its usefulness for SSN absorption measurement \cite{Bambrilla:2008}. In the last years, similar situation of CW pump has been considered demonstrating experimentally the SSN advantage in reconstructing the absorption profile of an object \cite{Moreau:2017}, and more recently in the construction of SSN raster scanning microscope\cite{Javier:2019}. Another important point is the effect of pump photon statistics on the generated twin photon statistics via SPDC. Intuitively, $0\leq\beta\leq 1$ accounts all of the last considered experimental situations. Therefore it is intriguing to consider the parameter $\beta$ in this model. Detected photon statistics and correlation are calculated using the above set of  transformation equations and vacuum state as follows:
\begin{eqnarray}
\langle N_{P}\rangle=\langle N_{R}\rangle=\eta\lambda,\nonumber
\end{eqnarray}
\begin{eqnarray}
\langle\Delta^{2}N_{P}\rangle=\langle\Delta^{2}N_{R}\rangle=\eta\lambda+\beta\eta^{2}\lambda^{2}
\end{eqnarray}
\begin{equation}
\langle\Delta(N_{P},N_{R})\rangle=\eta^{2}\left(\lambda+\beta\lambda^{2}\right),\nonumber
\end{equation}
\begin{eqnarray}
\sigma=\frac{\langle\Delta^{2}\left(N_{P}-N_{R}\right)\rangle}{\langle N_{P}\rangle+\langle N_{R}\rangle}=1-\eta,\nonumber
\end{eqnarray}
where $\langle N\rangle$ is the detected mean numbers of photons, the subscripts $P$ (1), $R$ (2) correspond to probe (signal), reference (idler) respectively, and $\langle\Delta^{2}N\rangle=\langle N^{2}\rangle-\langle N\rangle^{2}$, $\langle\Delta(N_{P},N_{R})\rangle=\langle N_{P}N_{R}\rangle-\langle N_{P}\rangle\langle N_{R}\rangle$ are the respective variance and covariance.
An important remark here is the fact that $\sigma$ is independent of $\beta$ and it is limited by the detection losses $\eta$. It implies the photon number correlation remain intact regardless of the values of $\beta$. Looking at the expression of variance, when $\beta=0$, $\langle\Delta N^{2}_{j}\rangle = \langle N_{j}\rangle$ (j=P,R), i.e,  the individual beam noise has Poissonian statistics, whereas for $\beta=1$, it is easy to see the noise of the individual beam has dominant thermal noise contribution. Thus, the model shows a way to switch from thermal to Poissonian statistics by varying the model parameter $\beta$ from one to zero.
\subsection{Symmetrical photon subtraction}

\begin{figure}[thb]
	\centering
	\includegraphics[width=7cm,height=8.5cm]{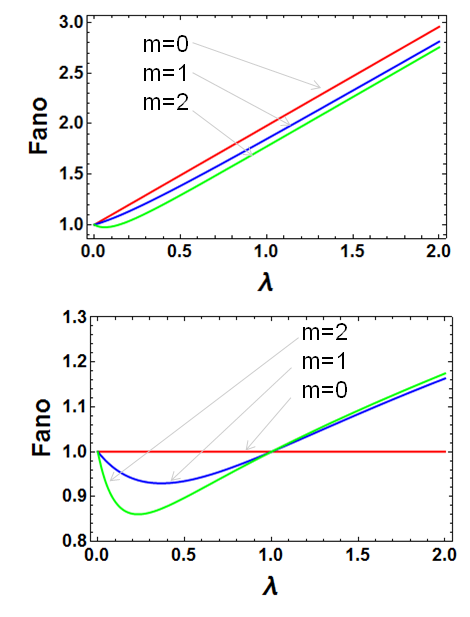}
	\caption{(Color Online) Plots of measured Fano Factor as a function of $\lambda$ with $\eta=0.98$ ($\gamma=0$) for different values of $m$: $m=0$ (solid red line), $m=1$ (solid blue line), and $m=2$ ( solid green line). We set $\beta=1$ (top) and  $\beta=0$ (bottom).}
	\label{fano}
\end{figure}

In this section, we shall see how to incorporate symmetrical photon subtraction taking into account the model parameter $\beta$. Theoretically photon subtraction is a non unitary operation, so a normalization factor is required for getting symmetrical photon subtracted squeezed vacuum state (thermal case)
\begin{equation}\label{SPATSV}
 \vert\Psi\rangle_{m}=N^{-}_{m}\left(\lambda\right)(\hat{a_{1}})^{m}(\hat{a_{2}})^{m} \hat{S_{1,2}}\left(\lambda\right)\vert 0,0\rangle_{1,2},
\end{equation}
where $N^{-}_{m}$ is the normalization constant of the form
\begin{equation}
N^{-}_{m}(\lambda)=m!(-i\sqrt{\lambda})^{m}P_{m}(i\sqrt{\lambda}),
\end{equation}
$P_{m}$ being the $m$th order Legendre's polynomial and $m$ is the number of subtracted photons. In an experimental scenario, two high transmittance beam splitter are placed in the paths of the two beams of the TWB. Two simultaneous clicks at the single photon detectors (SPDS) confirms the probabilistic generation of subtracted states as shown in the left image of fig \ref{equivalence}. Alternatively by injecting  m+1 component superposition state  \cite{Nigam:2019} to the NL in place of  vacuum, equivalently executes deterministically the  m photon subtraction operation as shown in the right image of fig \ref{equivalence}. Such superposition states can be experimentally generated \cite{Lee:2012}.Thus photon subtraction is incorporated by these $m+1$ component superposition state and the set of transformation equation defined in eq.\ref{c1} and eq. \ref{c2}. For subtracted states, statistics of the transformed operator  are obtained by using this input superposition  state in place of vacuum. We calculated the Fano factors of the photon subtracted states and they are plotted in the fig \ref{fano}.
It is evident that for $\beta=1$ (two individual beams are thermal for $m=0$), photon subtraction changes the statistics from thermal to sub-Poissonian for low values of the mean number of photon. On the other hand for $\beta=0$ (two individual beams are Poissonian as expected), photon subtraction further improves the statistics from Poissonian to sub-Poissonian and the improvement increases with the increasing $m$. This is an interesting result. Moreover in this case, the states have higher threshold in terms of $\lambda$ before they become thermal compared to the case of $\beta=1$. Our calculation shows that the noise reduction factor $\sigma$ remains the same regardless of the number of photon subtraction $m$. This is quite expected as for the balanced case, $\sigma$ is independent of the statistics of the state and only depends on the detection losses.
\subsection{Results}
Before interpreting the result, we would like to show the dependence of the mean number of photons per mode with the model parameter $\beta$ as $m$ changes from $0-2$.
\begin{figure}[thb]
	\centering
	\includegraphics[width=7.5cm,height=5.5cm]{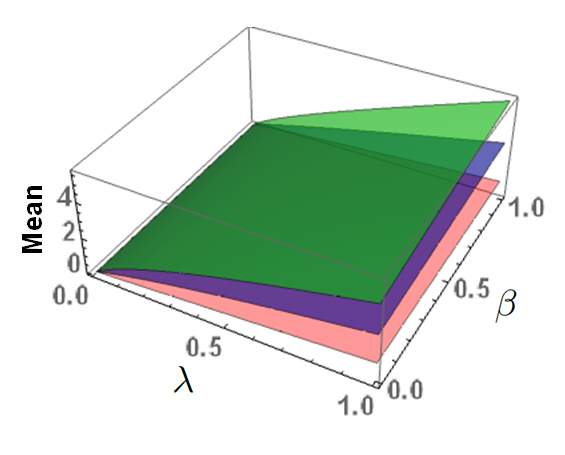}
	\caption{(Color Online) 3D plots of detected mean number of photons with $\eta=0.98$ gated by m detections: $m=0$ (bottom red sheet), $m=1$ (middle blue sheet), and $m=2$ (top green sheet).}
	\label{Meanc}
\end{figure}
Fig.\ref{Meanc} shows the non-linear rise of the mean number of photons with increasing $m$, and the  increment is monotonous with respect to $\beta$, i.e, minimum for $\beta=0$ and reaches maximum for $\beta=1$. Furthermore, for the last case, we checked in the low $\lambda$ limit, the rate of increment is maximum, i.e, four times for $m=1$ and nine times for $m=2$ with respect to $m=0$.
Substituting the expression of $F$ and  $\sigma$ in the uncertainty equations for different values of $m$ $(m=0-2)$, we worked out the uncertainties for above described three absorption estimators. The expressions are too cumbersome to present here, so we shall only depict the results graphically with relevant parameters of interest in the limiting cases. 
\subsubsection{Fixed squeezing parameter}
\begin{figure}[thb]
	\centering
	\includegraphics[width=7cm,height=13cm]{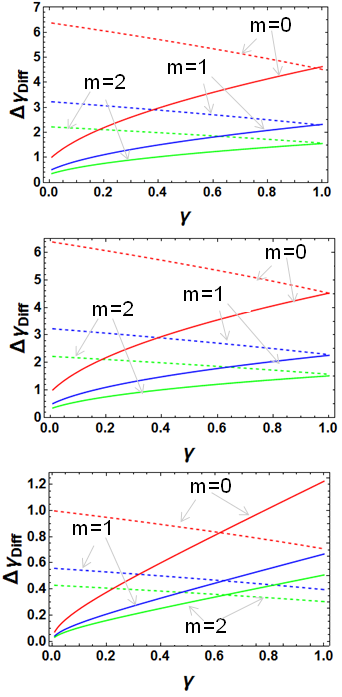}
	\caption{(Color online) Plots of Uncertainty in the number difference measurement versus absorption coefficient $\gamma$ with $\eta=0.98$ for different values of m: $m=0$ (solid red line), $m=1$ (solid blue line), and $m=2$ (solid green line). We set $\lambda=0.05$, $\beta=1$ (top), $\lambda=0.05$, $\beta=0$ (middle), and $\lambda=2$, $\beta=1$ (bottom).Dotted lines are the uncertainties evaluated using classical resources with average energies of $m$ photon subtracted state.}
	\label{Diffc}
\end{figure}

The analysis of comparing uncertainties for different $m$ at fixed squeezing has been carried at the same mean energy (photon number exposure) as the un-subtracted ($m=0$) state. In this way, r is fixed as mean energy $\lambda$ (for $m=0$)=$\sinh^{2} r$.
Uncertainties for number difference measurement is shown in fig \ref{Diffc}.
It shows SSN for different values of m (0-2) in this measurement. Albeit, photon subtraction show the advantage for all values of $\gamma$, particularly for low $\gamma$ and low $\lambda$, $m=2$ shows maximum advantage of almost three times over $m=0$. There is little change in the uncertainty as we vary $\beta$. As $\lambda$ increases, the thermal noise contribution in terms of $F$ in the uncertainty increases as per eq \ref{UncD}, as a result of which the SSN advantage for different m values is lost for relative high values of $\gamma$ as shown in figure. For the limit $\gamma\rightarrow 0$ (low absorption) and $\lambda\rightarrow 0$, the uncertainties in the number difference measurement for different m values scale as 
\begin{equation}
\Delta\gamma^{m=0}_{Diff}\approx\frac{\sqrt{2(1-\eta )}}{\sqrt{\eta\lambda}},\nonumber
\end{equation}
\begin{equation}
\Delta\gamma^{m=1}_{Diff}\approx\frac{\sqrt{(1-\eta )}}{\sqrt{2\eta\lambda}}
\end{equation}
\begin{equation}
\Delta\gamma^{m=2}_{Diff}\approx\frac{\sqrt{2(1-\eta )}}{3\sqrt{\eta\lambda}}.\nonumber
\end{equation}
On the other limiting case of complete absorption ($\gamma\rightarrow 1$) and for $\lambda\rightarrow 0$, the uncertainty scales as
\begin{equation}
\Delta\gamma^{m=0}_{Diff}\approx\frac{1}{\sqrt{\eta\lambda}},\nonumber
\end{equation}
\begin{equation}
\Delta\gamma^{m=1}_{Diff}\approx\frac{1}{2\sqrt{\eta\lambda}},
\end{equation}
\begin{equation}
\Delta\gamma^{m=2}_{Diff}\approx\frac{1}{3\sqrt{\eta\lambda}}.\nonumber
\end{equation}
This set of equation confirms the standard SNL for the limit $\gamma\rightarrow 1$ as per the discussion in \ref{AbsorptionDiff}. The improvement in the standard SNL for the limit $\gamma\rightarrow 0$ comes from a loss dependent factor $1-\eta$ in the numerator. The factor of improvements in the same limit due to different number of photon subtraction are also clear. \par
\begin{figure}[thb]
	\centering
	\includegraphics[width=7cm,height=13cm]{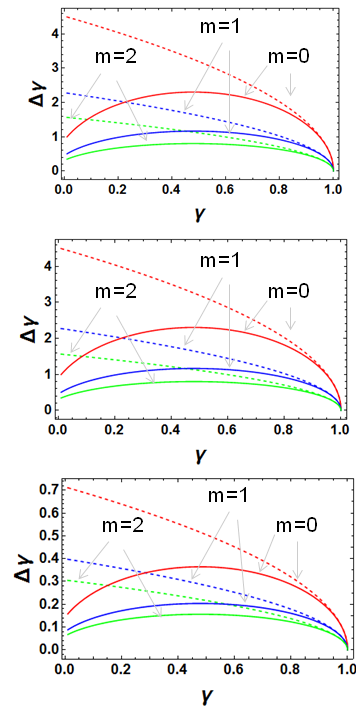}
	\caption{(Color online) Plots of uncertainty in the optimized balanced estimator versus absorption coefficient $\gamma$ with $\eta=0.98$ for different values of m: $m=0$ (solid red line), $m=1$ (solid blue line), and $m=2$ (solid green line). We set $\lambda=0.05$, $\beta=1$ (top), $\lambda=0.05$, $\beta=0$ (middle), and $\lambda=2$, $\beta=1$ (bottom). Dotted lines are the uncertainties evaluated using classical resources with average energies of $m$ photon subtracted state.}
	\label{Newc}
\end{figure}

 Normalized uncertainty for the optimized balanced absorption estimator is plotted in fig \ref{Newc}. 
Unlike the number difference estimator, regardless of the values of $\lambda$ for $\beta=1$ (thermal case), the uncertainty in the measurement for the optimized balanced absorption estimator is SSN enhanced for $\gamma<1$. This can be explained from the fact that the contribution of the thermal noise in the limit $\lambda\rightarrow\infty$ in the uncertainty of the measurement defined in eq \ref{newestimator} of \ref{AbsorptionNew} is less significant. We checked that the advantage of this optimized estimator is slightly better for $\beta=0$ (thermal case) compared to $\beta=1$ (Poissonian case) for both $\gamma\approx 1$, and for $\gamma\approx 0$. This advantage is further improved by the number of subtracted photon $m$ as shown in the figure. In the limiting case, the uncertainty of the new absorption estimator for different number of subtracted number $m$ can be expressed as follows:
for $\gamma\rightarrow 0$ (low absorption) and $\lambda\rightarrow 0$, the uncertainties in the new absorption estimator for different m values scale as 
\begin{equation}
\Delta\gamma^{m=0}_{Opt}\approx\frac{\sqrt{1-\eta^{2}}}{\sqrt{\eta\lambda}},\nonumber
\end{equation}
\begin{equation}\label{efficiencyc}
\Delta\gamma^{m=1}_{Opt}\approx\frac{\sqrt{1-\eta^{2}}}{2\sqrt{\eta\lambda}},
\end{equation}
\begin{equation}
\Delta\gamma^{m=2}_{Opt}\approx\frac{\sqrt{1-\eta^{2}}}{3\sqrt{\eta\lambda}}.\nonumber
\end{equation}
On the other limiting case of complete absorption ($\gamma\rightarrow 1$) and for $\lambda\rightarrow 0$, the uncertainty scales as
\begin{equation}
\Delta\gamma^{m=0}_{Opt}\approx\frac{\sqrt{1-\gamma}}{\sqrt{\eta\lambda}},\nonumber
\end{equation}
\begin{equation}
\Delta\gamma^{m=1}_{Opt}\approx\frac{\sqrt{1-\gamma}}{2\sqrt{\eta\lambda}},
\end{equation}
\begin{equation}
\Delta\gamma^{m=2}_{Opt}\approx\frac{\sqrt{1-\gamma}}{3\sqrt{\eta\lambda}}.\nonumber
\end{equation}
This set of equation says, unlike the number difference measurement, for $\gamma \rightarrow 1$, the uncertainty scales much better than the standard SNL and it improves further with the number of subtracted photon $m$. On the other side of the limit $\gamma\rightarrow 0$, apart from a factor $\sqrt{2}$, the improvement in the SNL compared to number difference measurement comes from a factor $1-\eta^{2}$ instead of $1-\eta$. Also in this case, there are factors of improvement due to photon subtraction, i.e $m+1$ times improvement in m photon subtraction.
\par
\begin{figure}[thb]
	\centering
	\includegraphics[width=7cm,height=13cm]{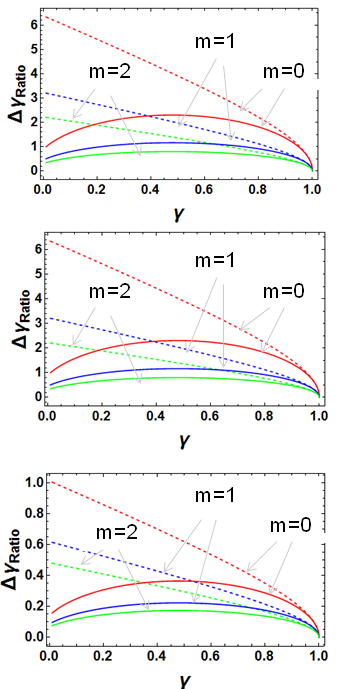}
	\caption{(Color online) Plots of uncertainty in the ratio measurement versus absorption coefficient $\gamma$ with $\eta=0.98$ for different values of m: $m=0$ (solid red line), $m=1$ (solid blue line), and $m=2$ (solid green line). We set $\lambda=0.05$, $\beta=1$ (top), $\lambda=0.05$, $\beta=0$ (middle), and $\lambda=2$, $\beta=1$ (bottom). Dotted lines are the uncertainties evaluated using classical resources with average energies of $m$ photon subtracted state.}
	\label{Ratioc}
\end{figure}

 A plot of uncertainty in the ratio measurement is shown in fig \ref{Ratioc}. 
It shows SSN limit for any values of mean number of photons per mode $\lambda$ as the uncertainty is unaffected by the thermal noise contribution and only relies on the photon number non-classical correlation as evident from eq \ref{Ratioestimator}. Again for fixed $\lambda$, we found the uncertainty for subtracted states (m=0-2) does not change much as $\beta$ changes from zero to one (maximum advantage is observed for $\beta=0$). Another notable thing is for fixed value of $\beta$, i.e zero, one or in-between, the uncertainty reduction is maximum for low values of $\lambda$, which is similar to the case for all our considered absorption estimators. We checked the uncertainty of the ratio measurement in the limiting case for different number of subtracted photons $m$ resembles to the uncertainty of optimized balanced estimator for $\lambda\rightarrow 0, \gamma\rightarrow 1$ except for $\lambda\rightarrow 0, \gamma\rightarrow 0$, where the uncertainty matches to the uncertainty of the number difference measurement. \par
\begin{figure}[thb]
	\centering
	\includegraphics[width=7cm,height=4.5cm]{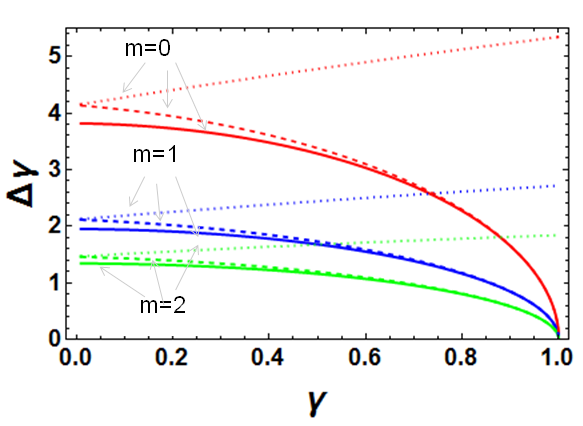}
	\caption{(Color online) Comparision of uncertainties among absorption estimators: number difference (dotted), ratio (dashed), and optimized balanced (Solid) versus $\gamma$ for $\eta=0.7$, $\beta=0$ and $\lambda=0.05$. Different colours correspond to different number of photon subtraction.}
	\label{compc}
\end{figure}

 We have plotted the uncertainties showing a comparison among them in fig \ref{compc}. 
The optimized estimator outperforms the number difference for the full range of absorption. Regardless of the values of $\gamma$, the optimized estimator and ratio performs equally well for low losses (high $\eta$). Nevertheless, in the limit of low $\gamma$ and high loss (low $\eta$) of about $30\%$, the optimized estimator performs better than ratio measurement because of the detection loss dependent factor $\sqrt{1-\eta^{2}}$ instead of $\sqrt{1-\eta}$ as per eq \ref{efficiencyc}. Photon subtraction shows  improvement of more than three  times compared to the case of previously considered $2\%$ of detection loss in all these three estimators. Furthermore, the overall magnitude of the uncertainty reduction for these three estimators decreases  at this high detection loss as they vary inversely with $\eta$ as evident from the uncertainty equations at their asymptotic limits. 
\subsubsection{Fixed per photon exposure}
Fixed per photon exposure analysis has been carried by balancing numerically the mean energies of subtracted states so that mean energies for different $m$ are equal to energy of ($m=0$) un-subtracted state. Since there is a $\beta$ dependence on mean energy, we further consider the energy balancing at a fixed model parameter. Before presenting the uncertainty result, we show the behaviour of Fano factor (F) in this energy balancing scenario.
\begin{figure}[thb]
	\centering
	\includegraphics[width=7cm,height=10cm]{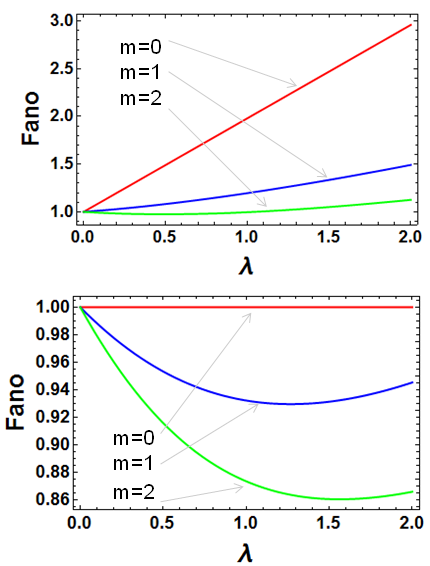}
	\caption{(Color online) Plots of Fano factor as a function of $\lambda$ with $\eta=0.98$ (and $\gamma=0$) for different values of $m$: $m=0$ (solid red line), $m=1$ (solid blue line), and $m=2$ ( solid green line). We set $\beta=1$ (top) and  $\beta=0$ (bottom).}
	\label{fanoeb}
\end{figure}	

For $\beta=1$ (thermal) case, $m=2$ remaining sub-poissonian until $\lambda=1$  while for $m=1$ and $m=0$ Fano factor remains super-Poissonian (fig \ref{fanoeb}). On the other hand, for $\beta=0$ (Poissonian case), sub-poissonian feature increases with $m$ other than $m=0$ and becoming maximal between $\lambda\simeq 1-2$. This shows a shift in $\lambda$ to higher number compared to the fixed squeezing case.
\begin{figure}[thb]
	\centering
	\includegraphics[width=7cm,height=10cm]{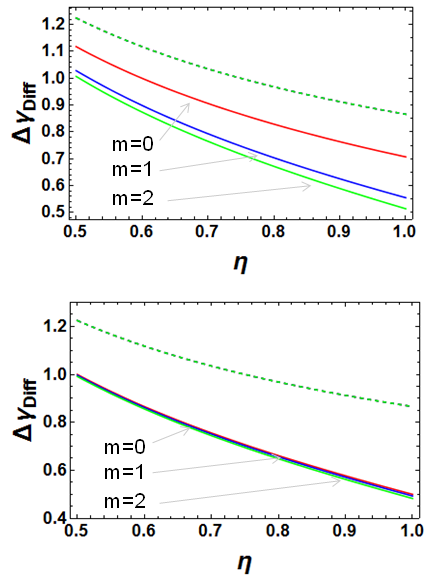}
	\caption{(Color online) Plots of uncertainty in the number difference estimator versus efficiency in fixed per photon exposure with $\gamma=0.5$ and $\lambda=2$ for different values of m: $m=0$ (solid red line), $m=1$ (solid blue line), and $m=2$ (solid green line). We set $\beta=1$ (top) and $\beta=0$ (bottom). Dotted lines are the SNL.}
	\label{diffceb}
\end{figure}

Uncertainties in the absorption coefficient $\gamma$ for number difference measurement is shown in the fig \ref{diffceb}. Unlike the result at fixed squeezing, we checked the advantage due to photon subtraction is almost lost in the regime of low $\lambda$ and $\gamma$ for $\beta=1$ (thermal). Nevertheless, some advantage still remains at relatively higher $\lambda$ and $\gamma$ compared to fixed squeezing. This advantage at high $\lambda$ can be related to F, and looking at the uncertainty expression in eq \ref{UncD}, the advantage at high $\gamma$ value can be understood from the fact that there must exist a value of $\gamma$  high enough ( loss sufficiently low) to reduce the uncertainty below the SNL. For instance, for value of $\gamma=0.5$ and $\eta\approx 1$, uncertainty using $m=0$ is below SNL and $m=2$ provides almost 20$\%$ advantage compared to $m=0$ although the advantage decreases at higher loss as evident from the fig \ref{diffceb} (top). At 50$\%$ of detection loss, $m=2$ provides almost 10$\%$ advantage compared to $m=0$. For $\beta=0$ (Poissonian) there is little advantage for $m=2$ due to limited improvement in F. The magnitude of uncertainty reduction for different $m$ comes closer to $m=2$ as shown in fig \ref{diffceb} (bottom). 
\begin{figure}[htb]
	\centering
	\includegraphics[width=7cm,height=10cm]{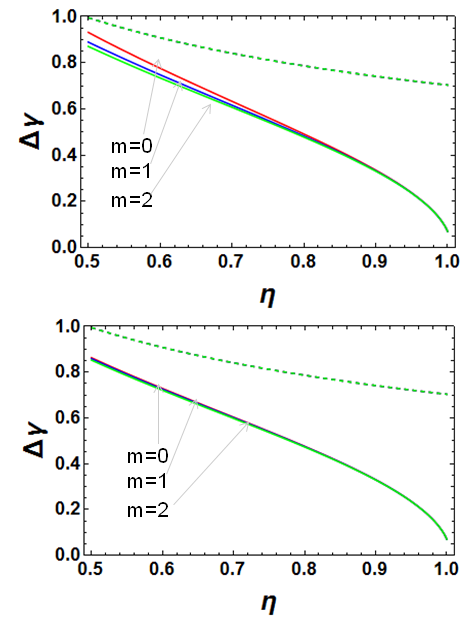}
	\caption{(Color online)  Plots of uncertainty in the optimized balanced estimator versus efficiency in fixed per photon exposure with $\gamma=0.01$ and $\lambda=2$ for different values of m: $m=0$ (solid red line), $m=1$ (solid blue line), and $m=2$ (solid green line). We set $\beta=1$ (top) and $\beta=0$ (bottom). Dotted lines are the SNL.}
	\label{paulceb}
\end{figure}
\par Uncertainty for optimized balanced estimator is plotted in fig \ref{paulceb}. Like the number difference estimator, the advantage is almost lost for low $\lambda$, but still photon subtraction gives a small advantage in the uncertainty reduction for high $\lambda$ and low $\gamma$, and the advantage is more at high detection losses unlike number difference measurement for $\beta=1$. For instance at 50$\%$ detection loss, about 10$\%$ advantage can be obtained for $m=2$ compared to $m=0$. This can be inferred from the uncertainty in eq \ref{newestimator} as the uncertainty reduction is more at high losses. The advantage due to different $m$ is very little and it is not very different compared to $m=0$ for $\beta=0$. Similar to number difference estimator, in this case, the magnitude of the uncertainty reduction for $m=0,1$ comes closer to $m=2$.
Since the uncerianty in ratio estimator in eq \ref{Ratioestimator} does not depend on the photon statistics, we check that for both  $\beta=1$ and  $\beta=0$, unlike the last two estimators, photon subtraction does not provide any advantage in the fixed per photon exposure to the absorption sample. 

 \section{Conclusions}\label{Conclusions}
 In summary, for the first time, we have successfully generated TBS whose individual beam photon statistics vary between thermal and Poissonian controlled by a 'modal averaging parameter' $\beta$, and demonstrated their usefulness for loss estimations. We have considered three different ways of measuring the absorption and found the best estimator among them in terms of measured uncertainty reduction. We established a clear connection between the uncertainty reduction for the three estimators with photon statics of individual mode of the TBS and their correlation by two factors namely Fano factor (F) and noise reduction factor ($\sigma$); non-classicality in these two factors allows sub-shot noise loss estimations. Furthermore, uncertainties from using number-difference and optimized balanced estimator depends on Fano factor and $\sigma$, while in the ratio measurement, it only depends on the mode correlation $\sigma$. \par We have then incorporated photon subtraction operation into the model which brings further improvement in photon statistics. Correlation remains the same and independent of  $\beta$, and for unit value of the model parameter only F changes from initial super-Poissonian to sub-poissonian with increase in the number of subtracted photons $m$ which further improves in magnitude for null value of the model parameter (initial statistics of the individual states are Poissonian before subtraction). Thus, we inferred that photon subtraction is advantageous for further  improving the individual photon statistics of correlated TBS with Poissonian statistics. \par All the improvements in photon statistics in terms of Fano factor are reflected in the measured uncertainties of the respective estimators. We have analysed them with respect to different number of subtracted photons $m$ by fixing both squeezing parameter and per photon exposure. At fixed squeezing, uncertainties in the three absorption estimators scale SSN and maximum uncertainty reduction advantage of three times is obtained for $m=2$ with respect to $m=0$ at very low absorption. The optimized balanced and ratio estimator outperform the number difference estimator. Although the ratio and optimized balanced estimators perform equally well at low detection losses, at higher losses the latter estimator is slightly better compared to the former. We notice no significant changes in the uncertainties for null and intermediate values of the model parameter. Furthermore we have explicitly computed the uncertainties at the asymptotic absorption limits ($\gamma=0,1$) which confirm all the discussed quantum enhancements in loss estimations due to photon subtraction and shows best absorption estimator as well. \par In the per photon exposure analysis, i.e when the average photons for different $m$ is held fixed before entering the sample, advantage due to photon subtraction in the first model almost subsides in the low $\gamma$ and low $\lambda$. Nevertheless, for high $\lambda$, some advantage of about $20\%$ is preserved for the number-difference at low detection losses, and nearly $10\%$ advantage is retained for the optimized balanced estimator at $50\%$ detection losses. Ratio measurement does not give any advantage in the context of photon subtraction at per photon exposure. Therefore, the improvement in uncertainty reduction due to photon subtraction in overall can be attributed to the improvement in mean number of photons and photon statistics. \par This work assumes balanced detection loss in two beams of the TBS, however for the unbalancing scenario, the improvement in correlation and statistics may vary and finding the best estimator becomes a potentially challenging task that we will address in future work.
\section{Acknowledgements}
The authors acknowledge support from EPSRC through the QUANTIC Hub EP/T00097X/1

\end{document}